\documentstyle[12pt]{article}

\def\rdots{\mathinner{\mkern1mu\raise1pt\vbox{\kern1pt\hbox{.}}\mkern2mu
   \raise4pt\hbox{.}\mkern2mu\raise7pt\hbox{.}\mkern1mu}}
\newcommand{\be}{\begin{equation}}
\newcommand{\ee}{\end{equation}}
\newcommand{\pa}{\partial}
\newcommand{\al}{\alpha}
\newcommand{\la}{\lambda}
\newcommand{\Z}{{\rm Z\kern-.35em Z}}
\newcommand{\bP}{{\rm I\kern-.15em P}}
\newcommand{\Q}{\kern.3em\rule{.07em}{.65em}\kern-.3em{\rm Q}}
\newcommand{\R}{{\rm I\kern-.15em R}}
\newcommand{\h}{{\rm I\kern-.15em H}}
\newcommand{\C}{\kern.3em\rule{.07em}{.65em}\kern-.3em{\rm C}}
\newcommand{\T}{{\rm T\kern-.35em T}}
\newcommand{\D}{{\kern-.5em }}

\begin{document}

\openup 1.5\jot

\centerline{Some Generalized BRS Transformations. \ II}
\centerline{ A Quantum Gravity Model}

\vspace{1in}
\centerline{Paul Federbush}
\centerline{Department of Mathematics}
\centerline{University of Michigan}
\centerline{Ann Arbor, MI 48109-1109}
\centerline{(pfed@math.lsa.umich.edu)}

\vspace{1in}

\centerline{Abstract}

Generalized BRS transformations such as introduced in Part I are applied to a model of quantum gravity.  This development is technically complex; but at the least should illustrate how much less rigid and more general of application are the new BRS transf
ormations.

\vfill\eject

\setcounter{equation}{22}

\noindent
III.  \underline{Formulation of the Generalized BRS Transformations in a Quantum Gravity Model.}

We turn to the model of quantum gravity presented in [5], with a covariant action in Euclidean space and the metric expressed non-linearly in terms of the real symmetric matrix $A_{\mu\nu}(x)$ by
\be
g_{\mu\nu} (x) = \left( e^{A(x)} \right)_{\mu\nu} .
\ee
We start from the gauge condition
\be	\pa_\mu A_{\mu\nu} (x) = 0     \ee
and then pass in the usual way to the use of a gauge fixing term (setting $\pa_\mu A_{\mu\nu} (x) = f_\nu(x)$ and then smearing over the $f_\nu$).  We write the action as 
\be	S = S_1 + S_2 + S_3	\ee
The basic action $S_1$ is covariant and given by
\be	S_1 = \int \sqrt{g} (\alpha R^2 + \beta R_{\mu\nu} R^{\mu\nu})	\ee
$S_2$ the gauge fixing term is as follows:
\be	S_2 = \frac \gamma 2 \int (\pa_\mu A_{\mu\nu}) (-\Delta) (\pa_{\mu '} A_{\mu' \nu})
\ee

Infinitesimal transformations of $g_{\mu \nu}$ are given by
\begin{eqnarray}
g_{\mu \nu} \rightarrow g_{\mu \nu} &+& g_{\alpha \nu} \pa_\mu \phi^\alpha + g_{\alpha \mu} \pa_\nu \phi^\alpha + (\pa_\alpha g_{\mu \nu})\phi^\alpha \\
&=& {\cal G}_{\mu\nu\al} \phi^\al
\end{eqnarray}
where $\cal G$ is defined by (28)-(29).  $S_1$ is invariant under this transformation.  Before we define the ghost action $S_3$ we need some notation.  We use quaternions as follows:
\begin{eqnarray}
\{\vec{\al}\} &=& \{\vec{i}, \vec{i}, \vec{k}\} \\
\pa_\mu &=& \pa_0 + \vec{i} \pa_x + \vec{j} \pa_y + \vec{k} \pa_z \\
\pa^{\vec{\al}}_\mu &=& \vec{\al} \times \pa_\mu \ .
\end{eqnarray}
Quaternion multiplication is understood in (32).  Thus
\be	\pa^{\vec{i}}_\mu = \pa_0 \vec{i} - \pa_x + \vec{k} \pa_y - \vec{j} \pa_z 	   \ee
and so (using (31) and (33))
\be	\pa^{\vec{i}}_0 = -\pa_x	\ee
\[	\pa^{\vec{i}}_1 = \pa_0	\]
\[	\pa^{\vec{i}}_2 = -\pa_z 	\]
\[	\pa^{\vec{i}}_3 = \pa_y \ . 	\]
Note
\be	\pa_\mu \pa^{\vec{i}}_\mu = 0 \ .	\ee
We introduce six real symmetric $3 \times 3$ matrices $M^i_{\vec{\al}\vec{\beta}} \ , i=1,...,6$.  And require
\be 	{\rm Tr} (M^i \; M^j) = \delta_{ij} \ .	\ee
The application of quaternions we are making is modelled from constructions in [7].

We now proceed to the development of the ghost action $S_3$, realizing the Fadeev-Popov determinant, {\it for a functional integral with respect to the variables} $A_{\mu\nu}$.  
The ghost fields will be $c^\alpha, \bar{c}_\al$, $\al = 0,...,3$ and
$d_i, \bar{d}_i, i=1,...,6$.  We define $F$ by
\be
\frac{\pa g_{\mu \nu}}{\pa A_{\mu' \nu'}} = \frac 1 2 
\left( \delta^{\mu'}_\mu \delta^{\nu'}_\nu + \delta^{\nu '}_\mu \delta^{\mu '}_\nu + F^{\mu ' \nu '}_{\mu \nu} \right)
\ee
and introduce  a number of differential operators
\begin{eqnarray}
{\cal L}^{(1)j}_{\mu \nu} &=& \left( \delta^{\mu '}_\mu \delta^{\nu '}_\nu + \delta^{\nu '}_\mu
  \delta^{\mu '}_\nu + F^{\mu ' \nu '}_{\mu \nu} \right) M^j_{\vec{\al}\vec{\beta}} \pa^{\vec{\al}}_{\mu '} \pa^{\vec{\beta}}_{\nu '} \\
\hat{{\cal L}}^{(2)j}_{\mu \nu} &=& \overleftarrow{\pa}^{\vec{\al}}_\mu \overleftarrow{\pa}^{\vec{\beta}}_\nu M^j_{\vec{\al}\vec{\beta}} \\
\hat{{\cal L}}^{(3)\beta}_{\mu \nu} &=& \overleftarrow{\pa}_\mu \delta^\beta_\nu +
\overleftarrow{\pa}_\nu  \delta^\beta_\mu \\
{\cal L}^{(4)}_{\mu \nu \beta} &=& g_{\beta \nu} \pa_\mu + g_{\beta \mu} \pa_\nu
+(\pa_\beta g_{\mu \nu}) = {\cal G}_{\mu \nu \beta}
\end{eqnarray}
In eq. (38) and (39) $j=1,...,6$ and in eq. (40) and (41) $\beta = 0,..,3$.  We also will
use
\begin{eqnarray}
{\cal L}^{(2)j}_{\mu \nu} &=& \pa^{\vec{\al}}_\mu \pa^{\vec{\beta}}_\nu M^j_{\vec{\al}\vec{\beta}} \\
{\cal L}^{(3)\beta}_{\mu \nu} &=& - \pa_\mu \delta^\beta_\nu - \pa_\nu \delta^\beta_\mu \ .
\end{eqnarray} 
>From the ${\cal L}^{(\ )}$ are constructed the following bilinear differential operators:
\begin{eqnarray}
t^{\beta '}_\beta &=& \sum_{\mu,\nu} \hat{{\cal L}}^{(3)\beta '}_{\mu \nu} {\cal L}^{(4)}_{\mu \nu \beta} \\
b^{ij} &=& \sum_{\mu,\nu} \hat{{\cal L}}^{(2)i}_{\mu \nu} {\cal L}^{(1)j}_{\mu \nu} \\
r^{\beta j} &=& \sum_{\mu,\nu} \hat{{\cal L}}^{(3)\beta}_{\mu \nu} {\cal L}^{(1)j}_{\mu \nu} \\
\ell^j_\beta &=& \sum_{\mu,\nu} \hat{{\cal L}}^{(2)j}_{\mu \nu} {\cal L}^{(4)}_{\mu \nu \beta} 
\end{eqnarray}
These enable us to write the ghost action conveniently:
\be	S_3 = \delta(T + R + L + B)		\ee
with:
\begin{eqnarray}
T &=& \int \bar{c}_{\beta '} t^{\beta '}_\beta c^\beta \\
B &=& \int \bar{d}_i b^{ij} d_j \\
R &=& \int \bar{c}_\beta  r^{\beta j} d_j \\
L &=& \int \bar{d}_j \ell^j_\beta  c^\beta 
\end{eqnarray}
It is sometimes helpful to form the bilinear operators in a block matrix
\be 
BL = \left( \begin{array}{cc}
t \ \ & \ \ r \\
 \\ 
\ell \ \  & \ \ b
  \end{array} \right)
\ee
We can studying (53) easily see that if $F^{\mu ' \nu '}_{\mu \nu}$ is set zero the ghost action is the same as the usual expression (i.e. if $A_{\mu \nu} = g_{\mu \nu}$ then the Fadeev-Popov determinant is the usual one).

The generalized BRS transformations may now be written down, generated by
\be	\phi^\al(x) = c^\al(x)\la + \int_y F^\al_\beta(x,y)c^\beta(y)\lambda	\ee
Here $F$ is an essentially arbitrary formal series in the $A_{\mu \nu}$.   One could add to eq. (54) a similar term to the last linear in $d_i(y)$; for simplicity we do not (it is straightforward to modify succeeding relations to accommodate this). The tr
ansformations follow:
\begin{eqnarray}
g_{\mu \nu}(x) &\rightarrow& g_{\mu \nu}(x) + {\cal G}_{\mu \nu \al}(x) \phi^\al(x) \\
\bar{d}_i(x) &\rightarrow& \bar{d}_i(x) + \left( \frac \gamma \delta \right) D_i(x) \la \\
\bar{c}_\al(x) &\rightarrow& \bar{c}_\al(x) + \left( - \frac 1 2 \frac \gamma \delta \right) 
(-\Delta) \pa_\mu A_{\mu \al}(x) \la + \frac \gamma \delta G_\al(x)\la \\
d_i(x) &\rightarrow& d_i(x) + \int_y \int_z W_{i\al j}(x,y,z) c^\al(y) d_j(z) \la \\ 
&\ \ \ \ \ \ & + \int_y \int_z X_{i\al \beta}(x,y,z) c^\al(y) c^\beta(z) \la \nonumber \\
c^\al(x) &\rightarrow& c^\al(x) + \left( \frac \pa {\pa x^\beta} c^\al(x)\right) c^\beta(x) \lambda 
\nonumber \\
& \ \ \ & + \int_y \int_z Z^\al_{\beta \gamma}(x, y, z) c^\beta(y) c^\gamma(z) \la \\
& \ \ \ & + \int_y \int_z Y^\al_{\beta i}(x, y, z) c^\beta(y) d_i (z) \la \nonumber
\end{eqnarray}
We will find it convenient to define $\tilde{F}$ by 
\be	\delta A_{\mu \nu} = \delta g_{\mu \nu} + \tilde{F}^{\mu ' \nu '}_{\mu \nu}  \delta g_{\mu' \nu'} 
\ee
where $\delta A$ and $\delta g$ are of course the changes in $A$ and $g$ under the BRS transformation.  $\tilde{F}$ is entirely determined from eq. (23) as
\be
\tilde{F}^{\mu ' \nu '}_{\mu \nu}(x) = \frac {\pa A_{\mu \nu}(x)} {\pa g_{\mu' \nu'}(x)} \frac 1 2  (1 + \delta_{\mu' \nu'}) - \delta^{\mu '}_\mu \delta^{\nu'}_\nu \ .
\ee
Similarly to the Yang-Mills case, the functions $D, G, W, X, Y, Z$ will have to satisfy certain relations specified in the next section.

\vspace{.25in}

\noindent
IV.  \underline{Relations ensuring $\Delta S - \Delta J = 0$.}

We use the notation of equations (12)-(14).  In the present case there are terms in $\Delta S - \Delta J $ of six types.  We require
\be	\Delta S_i - \Delta J_i = 0, \ \ \ \ \ i=1, \ldots, 6 \ \ .	\ee 
The six types of terms contain expressions of the following form, in an obvious notation:

1)	$d$

2)	$c$

3)	$\bar{d} \; c \; d$

4)	$\bar{c} \; c \; d$

5)	$\bar{c} \; c \; c$

6)	$\bar{d} \; c \; c$

\noindent
We proceed to list the required relations, each obtained by a tedious but direct computation.

\noindent
\underline{Condition 1}.  $\Delta S_1 - \Delta J_1 = 0$ provided
\[	2\gamma \Delta^2 D_i(x) + \gamma {\cal L}^{(2)i}_{rs} F^{rs}_{\mu \nu} {\cal L}^{(2)j}_{\mu \nu} D_j(x) - \int_z Y^\al_{\al i}(z, z, x)
\]
\be
 - \gamma {\cal L}^{(2)i}_{rs} F^{rs}_{\mu \nu} {\cal L}^{(3)t}_{\mu \nu} \left( G_t(x) + \frac 1 2 \Delta \pa_a A_{at}(x)\right) = 0 \ .
\ee

\noindent
\underline{Condition 2}.  $\Delta S_2 - \Delta J_2 = 0$ provided
\begin{eqnarray}
&-& \gamma \left( {\cal L}^{(3)\al}_{\mu \nu} G_\al(x) \right) \hat{{\cal G}}_{\mu \nu \beta} 
+ \gamma \left( {\cal L}^{(2)i}_{\mu \nu} D_i(x) \right) \hat{{\cal G}}_{\mu \nu \beta} \nonumber \\
&+& \int_z W_{i\beta i} (z,x,z) - \int_z \left( Z^\al_{\al \beta} (z,z,x) - Z^\al_{\beta \al}(z,x,z) \right) \nonumber \\
&-& \gamma \left[ (\Delta \pa_\mu \pa_a A_{as}) \tilde{ F}^{rt}_{\mu s}\right] \hat{{\cal G}}_{r t \beta} \nonumber \\
&-& \gamma \int_y \big(\Delta \pa_\mu \pa_a A_{as}\big)_y \left( {\cal G}_{\mu s \al} + \tilde{F}^{cd}_{\mu s} {\cal G}_{cd\al} \right)_y F^\al_\beta (y,x) \nonumber \\
&+& \frac \delta {\delta A_{\mu \nu}(x)} \left[ \left( \delta^r_\mu \delta^s_\nu + \tilde{F}^{rs}_{\mu \nu} \right) \hat{{\cal G}}_{r s \beta} \right] \nonumber \\
&+&\int_y \frac \delta {\delta A_{\mu \nu}(y)} \left[ \left( \delta^r_\mu \delta^s_\nu + \tilde{F}^{rs}_{\mu \nu}(y) \right) {\cal G}_{r s \alpha}(y) F^\al_\beta(y,x)  \right] =0 \ .
\end{eqnarray}
We have set
\be	
\hat{{\cal G}}_{\mu \nu \al} = -g_{\al \nu} \overleftarrow{\pa}_\mu - g_{\al \mu} \overleftarrow{\pa}_\nu + (\pa_\al g_{\mu \nu}).
\ee
In the second from last term in (64) the subscripts $y$ indicate the operators and functions are evaluated as functions and operators in variable $y$.  In the next to last term in (64) we wish the functional derivative to be evaluated before the derivativ
es in $\hat{{\cal G}}$.

In the remaining cases, $i \ge 3, \ \ \Delta J_i = 0$.   Before we proceed individually to these cases, it is convenient to work out some results that will be used in all these cases.  We study the effects of the BRS transformation (54)-(59) on $\cal G$ a
nd $F$.  We get
\be
F^{ab}_{\mu \nu} (x) \longrightarrow F^{ab}_{\mu \nu} (x) + \sum_{\mu' \ge \nu'} \frac {\pa F^{ab}_{\mu \nu} (x)} {\pa g_{\mu'\nu'}(x)} {\cal D}_{\mu ' \nu' \al}(x) \phi^\al(x)
\ee
\be = F^{ab}_{\mu \nu} (x) + P^{(1)ab}_{\mu \nu \al}(x) c^\al(x)\la + \int_y P^{(1)ab}_{\mu \nu \al}(x) F^\al_\beta(x,y) c^\beta(y)\la \ .
\ee
where $P^{(1)}$  is a first order differential operator defined by these equations.  In computing the change in $\cal G$ and the conditions to follow we may omit the first term on the right side of (54) and the second term on the right side of (59) as can
celling each other.  We then may write
\be
 {\cal D}_{\mu \nu \al}(x)  \longrightarrow   {\cal D}_{\mu \nu \al}(x) + \int_z \int_y \delta(x-y)
P^{(2)}_{\mu \nu \al \gamma}(x,y) F^\gamma_\beta (y,z) c^\beta(z)\la 
\ee
where $P^{(2)}$ is a differential operator first order in each variable defined implicitly by this equation (and straightforward to compute). 

\noindent
\underline{Condition 3}.  $\Delta S_3 = 0$ provided
\begin{eqnarray}
& &  {\cal L}^{(2)j}_{\mu \nu}(x) {\cal L}^{(1)k}_{\mu \nu}(x) W_{k \al i}(x,y,z) \nonumber \\
&+& {\cal L}^{(2)j}_{\mu \nu}(x) {\cal L}^{(4)}_{\mu \nu \beta}(x) Y^\beta_{ \al i}(x,y,z) \nonumber \\
&-& {\cal L}^{(2)j}_{\mu \nu}(x) \delta(y-x) \delta(z-x) P^{(1)ab}_{\mu \nu \al}(y) {\cal L}^{(2)i}_{a b}(z) \nonumber \\
&-& {\cal L}^{(2)j}_{\mu \nu}(x) \delta(z-x) P^{(1)ab}_{\mu \nu \beta}(x) F^\beta_\al(x,y)  {\cal L}^{(2)i}_{a b}(z) = 0 \ .
\end{eqnarray}

\noindent
\underline{Condition 4}.  $\Delta S_4 = 0$ provided
\begin{eqnarray}
& &  {\cal L}^{(3)\beta '}_{\mu \nu}(x) {\cal L}^{(4)}_{\mu \nu \beta}(x) Y^\beta_{\al i}(x,y,z) \nonumber \\
&+&  {\cal L}^{(3)\beta '}_{\mu \nu}(x) {\cal L}^{(1)j}_{\mu \nu }(x) W_{j \al i}(x,y,z) \nonumber \\
&-&  {\cal L}^{(3)\beta '}_{\mu \nu}(x) \delta(y-x)\delta(z-x)P^{(1)ab}_{\mu \nu \al}(y) {\cal L}^{(2)i}_{a b }(z)  \nonumber \\
&-&  {\cal L}^{(3)\beta '}_{\mu \nu}(x) \delta(z-x)P^{(1)ab}_{\mu \nu \beta}(x) F^\beta_\al(x,y) {\cal L}^{(2)i}_{a b }(z) = 0  \ .
\end{eqnarray}

\noindent
\underline{Condition 5}.  $\Delta S_5 = 0$ provided
\begin{eqnarray}
& &  {\cal L}^{(3)\beta }_{\mu \nu}(x) {\cal L}^{(4)}_{\mu \nu \beta'}(x) Z^{\beta '}_{\al \gamma}(x,y,z) \nonumber \\
&+&  {\cal L}^{(3)\beta }_{\mu \nu}(x) {\cal L}^{(1)j}_{\mu \nu }(x) X_{j \al \gamma}(x,y,z) \nonumber \\
&-&  {\cal L}^{(3)\beta }_{\mu \nu}(x) \delta(z-x) P^{(2)}_{\mu \nu \gamma \beta '}(z,x) F^{\beta '}_\al(x,y) = 0  \ .
\end{eqnarray}

\underline{Condition 6}.  $\Delta S_6 = 0$ provided
\begin{eqnarray}
& &  {\cal L}^{(2)i}_{\mu \nu}(x) {\cal L}^{(1)j}_{\mu \nu}(x) X_{j \al \beta}(x,y,z) \nonumber \\
&+& {\cal L}^{(2)i}_{\mu \nu}(x) {\cal L}^{(4)}_{\mu \nu \gamma}(x) Z^\gamma_{ \al \beta}(x,y,z) \nonumber \\
&-& {\cal L}^{(2)i}_{\mu \nu}(x) \delta(z-x)  P^{(2)}_{\mu \nu \beta \beta '}(z,x) F^{\beta '}_\al(x,y) = 0  \ .
\end{eqnarray}

$F$ is known, functions $D, G, W, X, Y, Z$ are solved for each as formal power series in the $A_{\mu,\nu}$.  Suppose all the functions are known through $n^{{\rm th}}$ order terms.  In the inductive procedure we then write the 
$(n+1)^{{\rm st}}$ order parts of the six condition equations.  The six resultant equations 
(each homogenous of degree $n+1$) contain respectively a single term in the$(n+1){{\rm st}}$ 
order part of $D, G, W, Y, Z, X$ the ``principal term" and all the rest of the terms in each equat
ion are expressible in terms of the terms of degree $\le n$ already known of the functions involved
 (and $F$).  The principal term arises as the first term in the condition equation, or part of the
 first term.  Thus for example in the first equation $\Delta^2 D_i$ gives rise to the principal term, 
and in the second equation the first term contains $(\delta_{\beta \al} \Delta + \pa_\beta \pa_\al)G_\al$ the principal term.  Our six equations in six unknown functions can thus be inductively solved as formal 
power series.  Convergence properties are not here studied.  The most important direction of further 
work is incorporation of these results into a renormalization scheme for the model.

\end{document}